\documentclass[aps,prl,twocolumn]{revtex4-2}
\usepackage{amsmath}
\usepackage{amssymb}
\usepackage{graphicx}
\usepackage{array}
\usepackage{dcolumn}
\usepackage{subfigure}
\usepackage{url}
\usepackage{color}
\usepackage{float}
\usepackage{multirow}
\usepackage{xr-hyper}
\usepackage[hidelinks=true]{hyperref}
\hypersetup{
  colorlinks   = true, %Colours links instead of ugly boxes
  urlcolor     = blue, %Colour for external hyperlinks
  linkcolor    = blue, %Colour of internal links
  citecolor   = red %Colour of citations
}

\begin{document}

\title{Structurally triggered orbital and charge orderings in TlMnO$_3$ and related compounds}
\author{Subhadeep Bandyopadhyay}
\email{subha.7491@gmail.com}
\author{Philippe Ghosez}
\email{Philippe.Ghosez@uliege.be}
\affiliation{Theoretical Materials Physics, Q-MAT, Université de Liège, B-4000 Sart-Tilman, Belgium}

\begin{abstract}
Rare earth perovskites ($R^{3+}$M$^{3+}$O$_3$), with   $e_g^1$ electronic occupation of the M $d$ states, display different types of metal-insulator transition.
 For manganites (M=Mn), metal-insulator transition is usually induced by the Jahn-Teller ($JT$) distortions, which stabilize orbital orderings (OO) at Mn sites. Among them, 
LaMnO$_3$ shows a $C$ type OO and crystallizes with $Pbnm$ structure. Whereas, TlMnO$_3$ shows a very distinct $G$ type OO  with an unusual $P\overline{1}$ structure. Employing first principles calculations, and symmetry mode analysis we rationalize structural and electronic origin of  $G$-type OO in TlMnO$_3$. Going further, we consider nickelates (M=Ni), where metal-insulator transition is driven by a breathing distortion, which stabilizes the charge ordering (CO) at Ni sites. Interestingly, different $JT$ and breathing distortions are very similar  MO$_6$ octahedral distortions  and stem from high frequency phonon modes of ideal $Pm\overline3m$ structure. Our comparative study reveals that following a common 
triggering mechanism these modes appear in their respective ground states.
  
\end{abstract}

\date{\today}

\maketitle
\begin{figure}[t]
\includegraphics[width=\columnwidth]{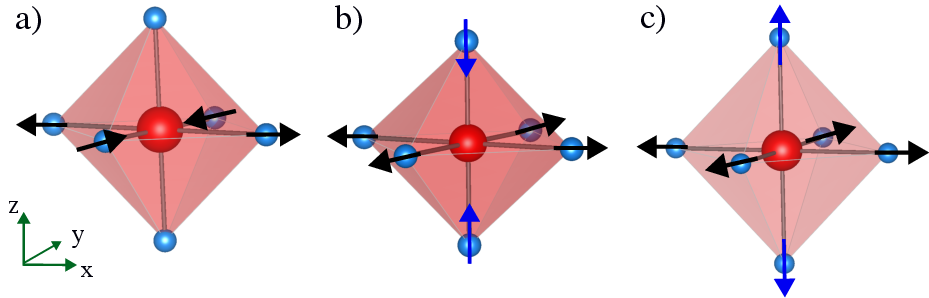}
\caption{Sketch of a (a) $Q_{2z}$, (b) $Q_{3z}$ and (c) $Q_B$ distortions of the MO$_6$ cages. According to canonical notations of Ref. \cite{Marcus_LMO}, $Q_{i\alpha}^J$ refers to a cooperative distortion type $i$, oriented along axis $\alpha$, with inter site modulation associated to $q$-point $J$.  }
\label{compare_modes}
\end{figure}

The $R^{3+}M^{3+}$O$_3$ perovskites with formal $e_g^1$ electron occupation of the metal $M$ $d$-states are well known to exhibit different kinds of metal-insulator transitions (MIT). Amongst them, LaMnO$_3$ and related lanthanide manganites ($Ln$MnO$_3$) crystallize  in a metallic $Pbnm$ phase at high temperature. The latter appears as a small distortion of the cubic reference perovskite structure and shows a pattern of oxygen octahedra rotations (OOR) labelled $a^-a^-c^+$ according to Glazer's notations \cite{Glazer}. Then, on cooling, $Ln$MnO$_3$ compounds keep their $Pbnm$ symmetry but exhibit a MIT related to orbital ordering (OO). More precisely in LaMnO$_3$, the MIT occurs at $T_{MIT}= 750$ K \cite{LMO_TM}. The bandgap opening arises from a $C$-type $d_{3x^2-r^2}$/$d_{3y^2-r^2}$ ordering of Mn $3d$ orbitals \cite{LMnO_OO} and is accompanied with the amplification of a $Q_{2z}^M$ cooperative Jahn-Teller (JT) distortion of the oxygen cages (following canonical notations of Ref \cite{Marcus_LMO}, see Fig.\ref{compare_modes}a), already allowed in the $Pbnm$ phase.  This takes place in the paramagnetic state while  $A$-type antiferromagnetic (AFM) ordering is appearing below $T_N = 140$ K \cite{LMO_TN}.

At odd with that, TlMnO$_3$ shows an unusual $P\overline{1}$ insulating ground state  \cite{TMO_ing}. The latter can also be seen as a small distortion of a $Pbnm$ metallic structure potentially stable at high temperature. But it arises here from a $G$-type  $d_{3x^2-r^2}$/$d_{3z^2-r^2}$ ordering of the Mn $3d$ orbitals and is accompanied with the appearance of a $Q_{2y}^R$ JT distortion \cite{TMO_PRB}. This distortion can be seen as a linear combination of $Q_{3z}^R$ and $Q_{2z}^R$ JT distortions (see Fig.\ref{compare_modes} and Supplementary Material (SM)note: S2), which can induce individually $G$-type OO with respectively  $d_{x^2-y^2}$/$d_{3z^2-r^2}$ and $d_{3x^2-r^2}$/$d_{3y^2-r^2}$ orderings at neighboring Mn sites. This OO appears again in the paramagnetic state and remains apparently present until decomposition temperature (820 K) \cite{TMO_ing}, while $E'$-type\footnote{Magnetic configuration of $E'$-type AFM is shown in SM note: S4, which includes ferromagnetic interaction between the Mn atoms within the [1 -1 0] plane and antiferromagnetic interaction  between these planes. Note that here $c$ is the long axis, whereas in the ref.\cite{TMO_PRB}, $b$ was chosen as the long axis.} magnetic ordering is found below $T_N = $ 92 K \cite{TMO_ing,TMO_PRB}.

A similar $G$-type OO is also  found in rare earth vanadates ($R$VO$_3$, $t_{2g}^2$ electron occupancy)
\cite{AVO_G_order1,AVO_G_order2,AVO_G_order3} but associated instead with a $Q_{2z}^R$ JT distortion, yielding a $Pbnm$ to $P2_1/b$  structural transition. This OO takes place in the paramagnetic state but at much lower temperature ($T_{MIT} \approx 150-200$ K) and gives rise to a $C$-type AFM ordering at low temperature ($T_N \approx 100$K). Some compounds with small $R$ cations, switch back to $Pbnm$ symmetry at low temperatures, but in an insulating phase combining $C$-type OO and $G$-type AFM ordering \cite{AVO}.

Then, TlNiO$_3$ \cite{TNO1,TNO2,TNO3} also possesses a  JT active cation Ni$^{3+}$, but shows a distinct $P2_1/n$ insulating ground state, identical to that of rare-earth nickelates ($R$NiO$_3$) \cite{TNO3}. Similarly to manganites, nickelates also show a $Pbnm$ metallic structure at high temperature and a MIT on cooling.  But, in this case, the gap opening is no more related to OO : it arises instead from a formal Ni$^{2+}$/Ni$^{4+}$ $G$-type charge ordering (CO) associated to a $Q_B^R$ breathing distortion of the oxygen cages (see Fig. \ref{compare_modes}c)\cite{Charge_vs_JT,YNO_Mercy}. These compounds show an AFM ordering that, depending of the cation size, can appear concomitantly with or at much lower temperature than the MIT.

Amazingly, the $Q_{2\alpha}^M$, $Q_{2\alpha}^R$, $Q_{3\alpha}^R$, $Q_{B}^R$ atomic distortions associated with distinct OO and CO are all very similar deformations of the oxygen octahedra. As such, they are {\it a priori} energetically costly and associated with phonons of very high frequencies in the dispersion curves of the reference cubic structure \cite{YNO_Mercy}. It is therefore natural to question (i) why these originally hard atomic motions want to condense at some point in those compounds and (ii) why distinct compounds selectively prefer to favor one of them over the others.      

\par

Here, we first focus on TlMnO$_3$ and rationalize how, starting from the cubic perovskite reference phase, it reaches its $P\bar{1}$ ground state going to an hypothetical intermediate $Pbnm$ phase and condensing a $Q_{2y}^R$ distortion.  Doing so, we reveal that $Q_{2\alpha}^M$, $Q_{2\alpha}^R$, $Q_{3\alpha
}^R$, $Q_{B}^R$ oxygen octahedra deformations are all triggered by OOR. We relate this first to their similar ability to open an electronic gap in the conduction bands and second in the similar role of OOR to tune O $2p$ - M $d$ hybridizations and produce this gap opening at the Fermi level, in a way similar to what was previously proposed for $R$NiO$_3$ compounds \cite{YNO_Mercy}. We then compare TlMnO$_3$ to LaMnO$_3$ and TlNiO$_3$. We show that the physics remains very similar in all three compounds, while their distinct final ground states result from subtle anharmonic couplings.

{\bf Computational details} -- 
Density functional theory calculations are performed relying on PBEsol \cite{PBESOL} exchange-correlation functional as implemented within   {\sc vasp} \cite{VASP1,VASP2,VASP3,VASP4}. We include Hubbard U and J correction to Mn-3$d$ orbital\cite{U} with  values of U, J = 8, 1 eV respectively, 
 which provide a good description of the lattice parameters and atomic distortions  of the ground state structures (see SM note: S3). Spin polarised calculations are performed with collinear magnetic arrangement. Following previous work of the nickelates \cite{YNO_Mercy}, we adopt a ferromagnetic (FM) spin ordering all along in our analysis, which   correctly capture the structural features without affecting our main conclusions.  
Symmetry adapted mode (SAM) analysis of the atomic distortions are performed using {\sc isodistort} \cite{Isodistort,Isodistort1}, taking the cubic $Pm\overline{3}m$ ABO$_3$ structure as reference. A suffix ($c$, $o$ or $m$) is typically added to mode labels, to specify to which phase they refer (cubic $Pm\overline3m$, orthorhombic $Pbnm$  or monoclinic $P2_1/n$).

\begin{figure}
\includegraphics[width=\columnwidth]{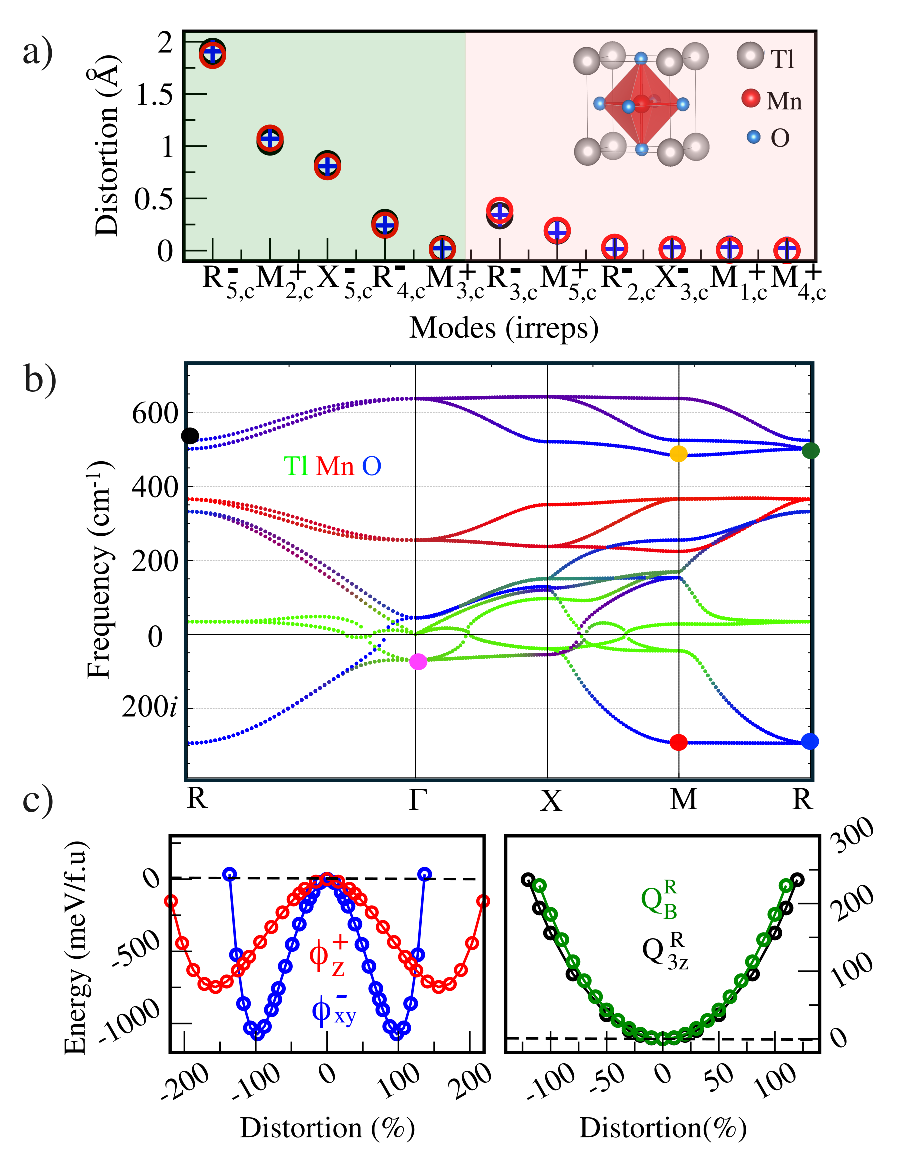}
\caption{(a) Symmetry-adapted mode decomposition of the full atomic distortion of $P\overline 1$
from the $Pm\bar{3}m$ reference (in the inset): 
comparison of the 
optimized structure (+) with experimental data of Ref. \cite{TMO_ing,TMO_PRB} (black and red open circles, respectively). Green and red backgrounds spread the distortions already allowed in $Pbnm$ from those further allowed in $P\bar{1}$. (b) Phonon dispersion curves in the $Pm\overline 3m$ structure of TlMnO$_3$. Unstable modes with imaginary frequencies appear below zero. Lines are colored according to the involvement of Tl (green), Mn (red) and O (blue) atoms in the eigendisplacement vector of each mode. Color dots locate some specific modes : $R_{5,c}^- \equiv \phi^-_{\alpha}$ (blue),  $M_{2,c}^+\equiv \phi^+_{\alpha}$ (red), $M_{3,c}^+ \equiv Q_{2\alpha}^M$ (orange), $R_{3,c}^- \equiv Q_{2\alpha}^R$/$Q_{3\alpha}^R$ (green), $R_{2,c}^- \equiv Q_B^R$ (black) and $\Gamma_4^-$ (magenta). (c) Energy wells corresponding to the condensation of distinct distortions in the $Pm\bar{3}m$ phase : (left) $\phi^-_{xy}$ (blue) and  $\phi^+_{z}$ (red);  (right) $Q_{3z}^R$ (black) and $Q_B^R$ (green). Amplitudes of $\phi^-_{xy}$, $\phi^+_{z}$ and $Q_{3z}^R$ distortions (relative to $Pm\overline3m$) are normalized to 1 (100\%) in the relaxed $P\overline1$ structure. Amplitude of $Q_B^R$ is taken the same as $Q_{3z}^R$. }
\label{wells}
\end{figure}

{\bf Ground state structure} -- 
We first optimize the $P \overline 1$ structure considering either FM or $A, ~C, ~G$ and $E'$ types AFM collinear orders (see SM note: S4). The magnetic moment is found to be of 4 $\mu_B$/Mn in all cases. In our calculations, the FM configuration appears to be the most stable, while the  $E'$-AFM order, reported as the experimental ground state \cite{TMO_ing}, is found 17 meV/f.u. higher in energy. This discrepancy might be an artifact of the used functional since previous hybrid functional calculations reported FM and $E'$-AFM configurations as nearly degenerated \cite{TMO_ing}. However, structural properties appear essentially independent of the spin ordering 
 and are well reproduced at the FM level (see Fig. \ref{wells}a) so that further analysis and discussion are restricted to the FM ordering. 

The $P\overline 1$ phase exhibits an insulating character. It shows a band gap  of  0.34 eV, originating due to the splitting of Mn-$e_g$ orbitals associated with JT distortions. 

The atomic structure of the $P\overline 1$ ground state can be characterized by decomposing the full atomic distortion respect to the cubic $Pm\bar{3}m$ parent phase in terms of SAM of distinct irreducible representations (see SM note: S5). As illustrated in Fig. \ref{wells}a, the fully relaxed structure is in excellent agreement with experimental data \cite{TMO_ing,TMO_PRB}. Amongst the dominant distortions, we can mention: $a^-a^-c^0$ OOR associated to $R_{5,c}^-$ SAM and further called $\phi^-_{xy}$; $a^0a^0c^+$ OOR associated to $M_{2,c}^+$ SAM and further called $\phi^+_z$; antipolar motions of Tl and O atoms along the $xy$ direction associated to $X_{5,c}^-$ SAM and further called $AP_{xy}$; JT distortions of the oxygen cages associated to $R_{3,c}^-(a,0)$ and $R_{3,c}^-(0,b)$ SAM (with $a>b$), corresponding respectively to previously introduced $Q_{3z}^R$ and $Q_{2z}^R$. Additional antipolar motions of Tl and O atoms ($R_{4,c}^-$ and $M_{5,c}^+$ SAM) are also present but play a more minor role. Then, further distortions are allowed by symmetry ($M_{3,c}^+ \equiv Q_{2z}^M$, $R_{2,c}^- \equiv Q_B^R$, $X_{3,c}^-$, $M_{1,c}^+$, $M_{4,c}^+$); although even more negligible, we will see later they are not necessarily irrelevant.

{\bf Proper structural instabilities} -- In line with its small Goldschmidt tolerance factor ($t=0.88$), TlMnO$_3$ exhibits two types of structural instabilities in the phonon dispersion curves of its reference cubic perovskite structure (Fig. \ref{wells}b). On the one hand, there is a flat line of phonon instabilities, associated to OOR, that is extending from $R_{5,c}^- \equiv \phi_{\alpha}^-$ ($\omega = 294i$ cm$^{-1}$) to $M_{2,c}^+ \equiv \phi_{\alpha}^+$ ($\omega = 293i$ cm$^{-1}$).  Consistently with that, the energy evolution related to the respective condensations of $\phi_{xy}^-$ ($a^-a^-c^0$ OOR pattern) and $\phi_z^+$ ($a^0a^0c^+$ OOR pattern) atomic distortions appear as typical double wells in Fig. \ref{wells}c. The joint condensation of $\phi_{xy}^-$ and $\phi_z^+$ in the cubic reference naturally brings the system to the common $Pbmn$ phase ($a^-a^-c^+$ OOR pattern). Although not observed experimentally ($P\bar{1}$ being stable until chemical decomposition \cite{TMO_ing}), this $Pbnm$ phase can be considered as the natural parent phase of the $P\bar{1}$ ground state since both show identical $a^-a^-c^+$ OOR pattern and dominant distortions (see Fig. \ref{wells}a). This metallic $Pbnm$ parent phase  appears 1.396 eV/f.u. lower in energy than the  $Pm\bar{3}m$ reference and 42 meV/f.u. higher in energy than $P\bar{1}$. 

On the other hand, in the cubic reference, there is also another line of phonon instabilities from $\Gamma_4^-$ ($\omega = 64i$ cm$^{-1}$)  to $X_5^-$ ($\omega = 55i$ cm$^{-1}$), related to polar and antipolar motions of the Tl atoms. As most often in perovskites with $t<1$, the $\Gamma_4^-$ polar instability -- which could make the system ferroelectric -- is latent but suppressed by the competition with more robust OOR \cite{Benede2013,ARCMP}. Then, we notice that the unstable character of the $X_5^-$ distortion contributes to stabilize the $Pbnm$ phase, in which it can naturally appear by symmetry \cite{ARCMP}.

Amazingly, although responsible for the $P\bar{1}$ ground state, $R_{3,c}^-$ modes ($\omega = 501$ cm$^{-1}$) related to $Q_{3\alpha}^R/Q_{2\alpha}^R$ JT distortions -- similarly to $M_{3,c}^+$ and $R_{2,c}^-$ modes related respectively to $Q_{2\alpha}^M$ or $Q_B^R$ distortions of the oxygen octahedra -- are very stiff and appearing at very high frequencies in the cubic phonon dispersion curves. Accordingly, the evolutions of the energy under condensation of $Q_{3z}^R$ and $Q_B^R$ atomic distortions correspond to typical single wells in Fig. \ref{wells}c. This then questions the appearance of $R_{3,c}^-$ modes in the $P\bar{1}$ ground state (Fig. \ref{wells}a), which cannot be considered as proper.

{\bf Triggered mechanism} -- 
\begin{figure}[t]
\includegraphics[width=\columnwidth]{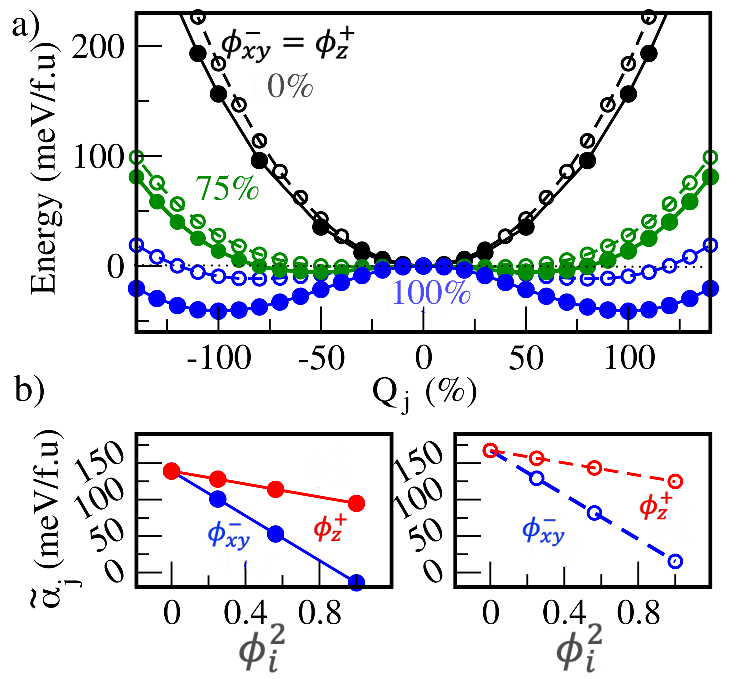}
\caption{Triggered mechanism. (a) Evolution of the energy $E$ in terms of the amplitude of Jahn-Teller distortion $Q_{3z}^R$ (solid line and filled circles) and breathing distortion $Q_B^R$ (dashed line and open circle) for fixed amplitudes of oxygen rotations ($\phi_{xy}^- = \phi_{z}^+$ from 0 to 100\%) in the FM cubic cell of TlMnO$_3$ (at a volume similar to the experimental ground state \cite{TMO_ing}). Curves have been shifted to aligned at the origin.  (b) Linear evolution of the energy curvature at the origin, $\tilde{\alpha_j}$ along $Q_j = Q_{3z}^R$ (left) and $Q_j = Q_{B}^R$ (right), in terms of the square of the amplitude of the individual distortions $\phi_{xy}^-$ (blue) and $\phi_{z}^+$ (red). Amplitude of distortions are normalized as in Fig. \ref{wells}.}
\label{triger}
\end{figure}
The intriguing condensation  of originally hard distortions in TlMnO$_3$ appears very similar to what was previously reported for YNiO$_3$ \cite{YNO_Mercy} and can be rationalized from anharmonic distortion couplings. Expansion of the Born-Oppenheimer energy around the cubic reference phase in terms of OOR amplitudes $\phi^-_{xy}$ and $\phi^+_{z}$ and a given oxygen octahedra distortion amplitude $Q_j$ can be written as :
\begin{equation}
\begin{aligned}
E ~\propto~ &\alpha_\mathrm{xy}{\phi^-_\mathrm{xy}}^2 + \beta_\mathrm{xy}{\phi^-_\mathrm{xy}}^4 + \alpha_\mathrm{z} {\phi^+_\mathrm{z}}^2 +
\beta_\mathrm{z} {\phi^+_\mathrm{z}}^4 \\ 
& + \lambda_\mathrm{xy,z}{\phi^-_\mathrm{xy}}^2{\phi^+_\mathrm{z}}^2 +\alpha_{j} Q_{j}^2 + \beta_{j} Q_{j}^4 \\
& + \lambda_{\mathrm{xy,j}}{\phi^-_\mathrm{xy}}^2Q_{j}^2 + \lambda_{\mathrm{z,j}}{\phi^+_\mathrm{z}}^2Q_{j}^2,
\end{aligned}
\label{equation1}
\end{equation}
in which $\alpha_j$ describes the energy curvature respect to $Q_j$ in the cubic phase and is proportional to $\omega_i^2$ while $\lambda_{ij}$ quantifies the bi-quadratic coupling between distortions $i$ and $j$. Fitted values of the parameters on DFT data are reported in SM, Table. S4.

According to the biquadratic couplings in Eq. (1), finite amplitudes of the OOR renormalize the original energy curvature  $\alpha_j$ respect to $Q_j$ :
\begin{equation}
\begin{aligned}
\tilde \alpha_{j} ~=~ \frac{\partial^2 E}{\partial Q_j^2} ~=~ \alpha_{j} +\lambda_{\mathrm{xy,j}}{\phi^-_\mathrm{xy}}^2 + \lambda_{\mathrm{z,j}}{\phi^+_\mathrm{z}}^2
\end{aligned}
\label{equation2}
\end{equation}
and can soften the mode in case $\lambda_{ij}<0$ ({\it cooperative} bi-quadratic coupling).
Fig. \ref{triger}(a) highlights the crucial role of OOR in inducing the $Q_{3z}^R$ distortion through such a mechanism: while $Q_{3z}^R$ is originally associated to a single energy well in the cubic phase (Fig. \ref{wells}c), the joint appearance of $\phi^-_{xy}$ and $\phi^+_{z}$ in the $Pbnm$ phase progressively turns it into a double energy well. Fig. \ref{triger}(b) points out that the renormalization of the curvature at the origin ($\tilde{\alpha}_j$) arises from cooperative bi-quadratic couplings of $Q_{3z}^R$ with OOR (i.e. $\lambda_{xy,j} < 0$ and $\lambda_{z,j}  < 0 $ in Eq. 1-2). This demonstrates that the appearance of $Q_{3z}^R$ is not {\it proper} but conditioned to the appearance and amplitude of $\phi^-_{xy}$ and $\phi^+_{z}$.  Following Holakovsky \cite{Holakovsky}, we refer to such a softening of a hard distortion from the cooperative biquadratic coupling with another primary order parameter as a {\it triggered mechanism} \cite{YNO_Mercy,Zhang-24,ARCMP}\footnote{Originally \cite{Holakovsky}, the name {\it triggered transition} was introduced to describe a situations in which a negative biquadratic coupling between first and secondary order parameters is present in the energy expansion and large enough for all of them to condense simultaneously at a unique transition. By extension, following Ref. \cite{YNO_Mercy,Zhang-24,ARCMP}, we speak here of a {\it triggered mechanism} when a negative biquadratic coupling is present in the Landau expansion. Notice that TlMnO$_3$ remains $P\bar{1}$ until decomposition temperature \cite{TMO_ing} so that the exact sequence of phase transition remains hypothetical.}. Highlighting this mechanism is not only of academic interest since, as demonstrated in nickelates \cite{Liao-18}, it provides a concrete pathway to tune the MIT from the control of OOR.

As illustrated in Fig. \ref{triger}, not only $Q_{3z}^R$ but also $Q_B^R$ is triggered by the OOR and it appears to be similarly the case for $Q_{2z}^M$ and $Q_{2z}^R$ (see SM note: S7 and S8).  This highlights that, in fact, all JT and breathing oxygen octahedra deformations are triggered by OOR.

{\bf Origin of the triggered mechanism} -- In YNiO$_3$, it was shown that the triggering of the breathing distortions by OOR is intimately linked to the electronic properties \cite{YNO_Mercy}.
\begin{figure}[t]
\includegraphics[width=\columnwidth]{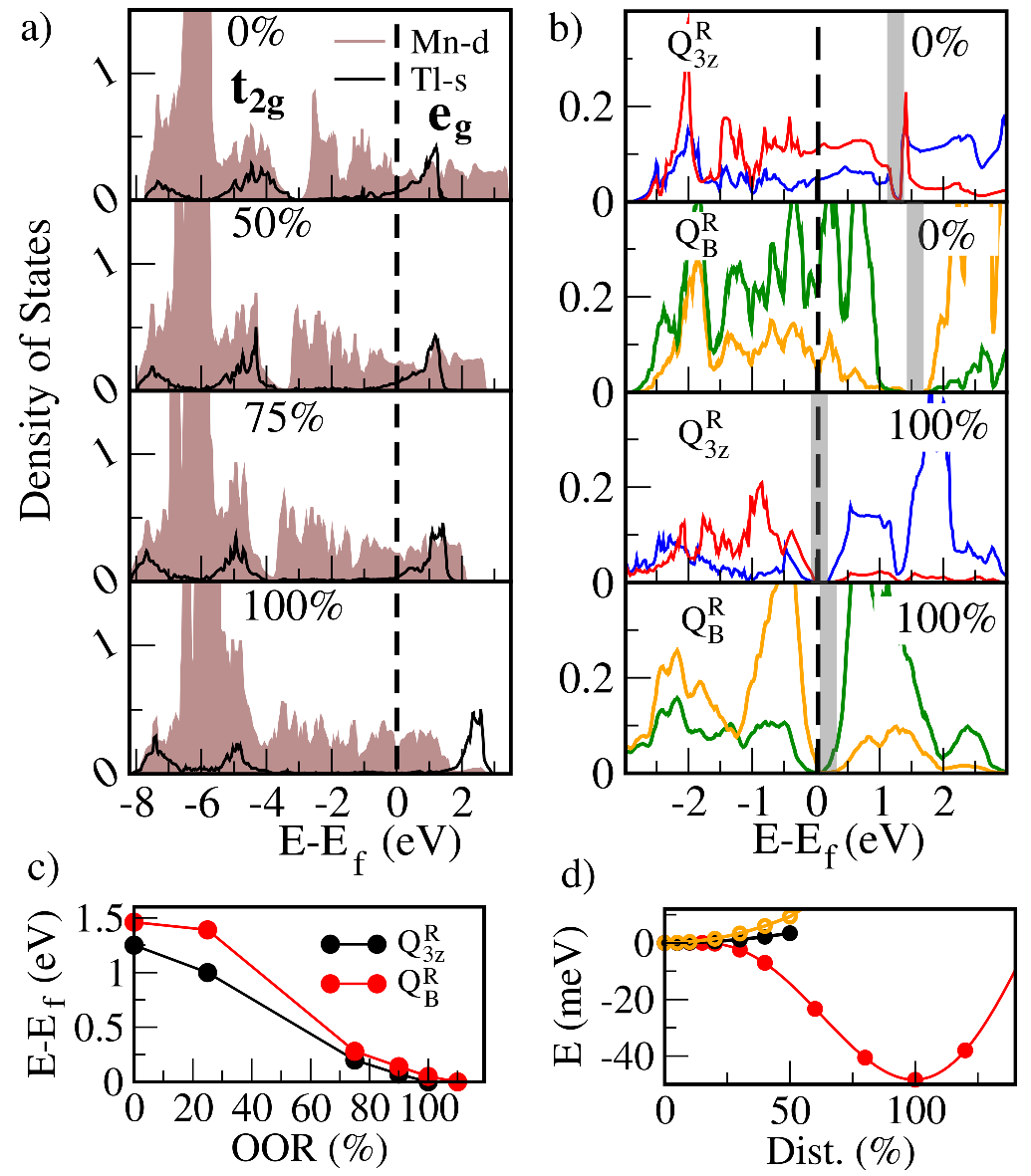}
\caption{(a) Partial DOS of Mn $3d$ and Tl $6s$  spin-up states at different joint amplitude of OOR $\phi_{xy}^-$ and $\phi_z^+$. (b) Evolution of the DOS under condensation of Q$^R_{3z}$ and Q$_B$ at 0 \% of OOR (top two panels) and 100 \% of joint OOR (botom two panels). Red and blue lines refer to  Mn-$d_{x^2-y^2}$ and Mn-$d_{3z^2-r^2}$ states while yellow and green lines refer to  $e_g$ states for respectively ~Mn$^{3+\delta}$ and ~Mn$^{3-\delta}$ atoms.
Black dashed line shows the position of $E_f$, and grey shaded area highlights the formation of the gap. (c) Evolution of the energy at which gap opens under condensation of Q$^R_{3z}$ (black) and Q$_B^R$ (red) as a function of the amplitude of joint OOR. (d) Energy/4 f.u. under condensation of  $\Gamma_{2,m}^+$ distortion (black), $\Gamma_{1,m}^+$+$\Gamma_{2,m}^+$ distortion (orange) and $\Gamma_{1,m}^+$+$\Gamma_{2,m}^+$ with further relaxation of  symmetry allowed strain (red).}
\label{JT}
%\end{figure}
\end{figure}
This view might be generic and apply to TlMnO$_3$ but needs further confirmation. 

In the cubic phase of TlMnO$_3$, the $t_{2g}^3e_g^1$ electronic configuration of the Mn$^{3+}$ gives rise to a high-spin arrangement. As shown in Fig. \ref{JT}a, this leads to metallicity, with partially occupied spin-up $e_g$ states at the Fermi energy ($E_f$). The occupied Mn-$t_{2g}$ states are separated due to octahedral crystal field and lie below -4 eV. 
Then, anti-bonding Tl$6s$-O$2p$ states are also present at $E_f$. The condensation of OOR (here jointly $\phi_{xy}^-$ and $\phi_z^+$
)  in the cubic phase, on the one hand, progressively decreases the hybridizations between Mn$3d$ and O$2p$ states, which is narrowing the bandwidth of the Mn-$d$ states (Fig. \ref{JT}a). 
and, on other hand, gradually increases the hybridizations between Tl$6s$ and O$2p$, which  enhances the energy splitting between bonding and anti-bonding states.

The condensation of $Q_{3z}^R$ JT distortions into the cubic phase (Fig. \ref{JT}b) splits energy levels of Mn-$d_{x^2-y^2}$ and $d_{3z^2-r^2}$ states, and opens a gap within the Mn-$e_g$ states. In this cubic phase, this gap opening nevertheless happens at $E_{g}({Q_{3z}^R}) = 1.25$ eV above $E_f$, but the system remains metallic. As previously discussed, the activation of OOR  reduce the Mn$3d$-O$2p$ states hybridizations, narrowing the bandwidth. Accordingly, as illustrated in Fig. \ref{JT}c, the condensation of OOR progressively move $E_{g}({Q_{3z}^R})$ toward $E_f$. The  $Q_{3z}^R$ JT mode becomes unstable at 95\% of OOR, when $E_{g}({Q_{3z}^R}) \approx E_f$ (i.e. when the gap opening produced by the activation of $Q_{3z}^R$ start lowering occupied states at $E_f$). As such, and similarly to what was previously reported in nickelates \cite{YNO_Mercy}, the MIT transition can be seen as a Peierls instability, which is however not present in the cubic reference phase but has been triggered by the OOR distortions in the $Pbnm$ phase.

This is not restricted to $Q_{3z}^R$ and a very similar picture applies to the breathing distortions $Q_B^R$.  When $Q_B^R$ distortions (with same amplitude as $Q^R_{3z}$) are forced into the cubic phase, charge ordering (CO) occurs  at the Mn sites, yielding inequivalent  $\sim$Mn$^{3+\delta}$ and $\sim$Mn$^{3-\delta}$ charge states. This can be identified from the occupancy of the $e_g$ states (see in Fig. \ref{JT}b) and is able to open a gap at $E_{g}({Q_B^R}) = 1.4$ eV above $E_f$. Together with the condensation of OOR (Fig. \ref{JT}c), $E_{g}({Q_B^R})$ gradually evolves toward $E_f$,  akin to what we  see for $Q_{3z}^R$. Consistently with the fact that $Q_{B}^R$  mode requires larger amplitude of OOR then $Q_{3z}^R$ to become unstable,  $E_{g}({Q_B^R})$ reaches $E_f$ at larger OOR amplitudes.
A similar behavior is also observed for  $Q_{2z}^R$ and  $Q_{2z}^M$, showing that the mechanism is generic to all types of oxygen octahedra distortions.

{\bf Path to the ground state} -- 
Although $Q^R_{3z}$, $Q^R_{2z}$, $Q^M_{2z}$ and $Q_B^R$ modes are all triggered by $\phi_{xy}^-$ and $\phi_z^+$ OOR, only $Q_{3z}^R$ appears to be  unstable in the relaxed $Pbnm$ phase. This can be explained from the fact that this phase does not only condense OOR ($R_{5,c}^-$ and $M_{2,c}^+$ in Fig. \ref{wells}a) but can also accommodate by symmetry $Q_{2z}^M$ distortions ($M_{3,c}^+$ in Fig. \ref{wells}a), as also seen in other compounds \cite{Varignon_JT,AVO}. Although the amplitude of $Q_{2z}^M$ remain small in $Pbnm$, this distortion competes with all other oxygen octahedra distortions through positive biquadratic couplings. As a result, the instabilities of $Q_{2z}^R$ and $Q_B^R$ triggered by the OOR are suppressed while only the $Q_{3z}^R$ mode remains slightly unstable in $Pbnm$ (see SM Table. S5), consistently with what has been discussed before. 
Condensing this unstable mode produces a structural transition from $Pbnm$ to $P2_1/n$, inducing a $G$-type OO of the Mn-$d_{3z^2-r^2}$/$d_{x^2-y^2}$ orbitals as depicted in SM Fig. S4. This $P2_1/n$ phase is insulating with a band gap of 0.31 eV. $P2_1/n$ appears 12 meV/f.u higher in energy than the $P\overline1$ ground state but does not show any dynamical instability (i.e. it appears as a local minimum).

To reach the $P\overline 1$ ground state from this $P2_1/n$ intermediate phase requires the  condensation of $Q_{2z}^R$  distortions. The latter is dominantly associated to  a $\Gamma_{2,m}^+$, stable phonon mode of $P2_1/n$, lying at 80 cm$^{-1}$. Condensation of this $\Gamma_{2,m}^+$ mode properly brings the system to $P\overline 1$ and is compatible with $\Gamma_{1,m}^+$ mode relaxation and the appearance of related $\eta_{2,m}^+$, $\eta_{1,m}^+$ strains in the optimized $P \overline1$ structure.  As shown in Fig. \ref{JT}d, the single energy well of the $\Gamma_{2,m}^+$  becomes even slightly stiffer on inclusion of additional  $\Gamma_{1,m}^+$. The energy curvature becomes negative only when  strain relaxation is considered, indicating the crucial role of strain-phonon coupling to induce the structural transition from $P2_1/n$ to the $P\overline 1$ structure. This transition then changes the type of JT distortion from $Q_{3z}^R$ to $Q_{2y}^R$ (which appears as a combination of original $Q_{3z}^R$ and additional $Q_{2z}^R$, see SM note S2), then changing as well the orbital ordering to Mn-$d_{3x^2-r^2}$/$d_{3z^2-r^2}$. 

Interestingly, decomposition of the $\Gamma_{2,m}^+$ shows contribution of the cubic $M_{1,c}^+$ distortion, which produces breathing like distortion of TlO$_{12}$ polyhedra, as reported recently for BiNiO$_3$\cite{BNO_Subhadeep}. However the effect is very weak in TlMnO$_3$. All this points out that the $P\overline 1$ ground state was not granted and arises from very subtle anharmonic interactions between different modes and strains.
 
{\bf Related compounds} --
 
Interestingly, the triggering of JT and breathing distortions by OOR is not restricted to TlMnO$_3$ but similarly appears for instance in other related compounds like LaMnO$_3$ and TlNiO$_3$ (see SM note: S7), making the mechanism generic. The fact that all these three compounds show distinct ground-state can again be rationalized from the strength of anharmonic interactions (see SM Table. S5). In LaMnO$_3$, the $Q_{2z}^M$ mode inherent to the $Pbnm$ phase suppresses the other instabilities triggered by the OOR so that the system stays in the $Pbnm$ phase, stabilizing a $C$-type OO. In TlNiO$_3$, the $Q_{2z}^M$ mode does not fully suppress the instability associated to $Q_B^R$ in $Pbnm$, so that the system evolves to a $P2_1/n$ ground state showing G-type charge ordering at Ni sites.  This clarifies that all these compounds are more similar than there distinct ground states might suggest and that OOR plays a key role in inducing the MIT. 

{\bf Conclusions} --
Our study highlights a very generic triggering of all oxygen octahedra deformations (JT and breathing) by OOR in $e_g^1$ perovskites, while distinct specific ground states may arise from subtle anharmonic distortion couplings. This points out and rationalizes that the electronic ground state and properties in these compounds might be controlled from the manipulation of these anharmonic couplings. Supporting this idea, biaxial strain induced  OO in HoNiO$_3$\cite{HoNiO3_hexu} or epitaxial strain induced stabilization of $Q_{3z}^R$ JT distortion in LaMnO$_3$\cite{Marcus_LMO}  have been previosuly reported. Similar tuning strategy has also been suggested for alkaline-earth ferrite perovskites, where a control over alternative charge and orbital orderings was predicted under tensile strain \cite{AFeO3_zhang}.
\\

%\begin{figure}[h]
%\includegraphics[width=\columnwidth]{totaldos.eps}
%\caption{}
%\label{triger}
%\end{figure}
$Acknowledgement:$ SB thanks He Xu for useful discussions and technical support. This work was supported  by F.R.S.-FNRS Belgium under PDR grant T.0107.20 (PROMOSPAN). The authors acknowledge use of the CECI supercomputer facilities funded by the F.R.S-FNRS (Grant No. 2.5020.1) and of the Tier-1 supercomputer of the Fédération Wallonie-Bruxelles funded by the Walloon Region (Grant No. 1117545).

\bibliographystyle{apsrev4-2} 
\bibliography{main.bib}

\end{document}